\documentclass[11pt,a4paper]{article}

\RequirePackage[l2tabu, orthodox]{nag}

\usepackage{jheppub}
\usepackage{amsthm,amssymb,amsmath,epic,float}
\usepackage{rotating,epsfig,indentfirst,array,varioref}
\usepackage{appendix,tikz,mathtools}
\usepackage{setspace}
\usepackage{tikz}
\usetikzlibrary{decorations.pathreplacing}
\usetikzlibrary{calc}
\usepackage{enumitem}
\usepackage{hyphenat}
\usetikzlibrary{positioning}
\usepackage{graphicx}  
\usepackage{adjustbox}

\usepackage{arydshln}

\def\bbt{\bibitem}
\def\be{\begin{equation}}
\def\en{\end{equation}}
\def\ber{\begin{eqnarray}}
\def\enr{\end{eqnarray}}
\def\nmb{ \nonumber\\}
\def\d{\partial}

\def\al{\alpha}

\def\im{\imath}

\def\lm{\lambda}
\def\Lm{\Lambda}

\def\om{\omega}

\def\eps{\epsilon}

\def\bk{{\bf k}}

\def\fr12{\frac{1}{2}}

\usepackage{jheppub}
\makeatletter
\def\@fpheader{\vspace{-.1cm}}
\makeatother

\title{\boldmath  Conformal bootstrap and Heterotic string Gepner models\\}

\author{Alexander Belavin,}
\author{Sergej Parkhomenko}

\affiliation{Landau Institute for Theoretical Physics, 142432 Chernogolovka, Russia}

\emailAdd{belavin@itp.ac.ru}
\emailAdd{sergej.e.parkhomenko@gmail.com}

\abstract{In Gepner's pioneering work, the requirement that leads to a model having the desired $N=1$ Spacetime supersymmetry and 
$E(8)\times E(6)$ Gauge symmetry was the requirement that the spacetime symmetry is compatible with modular invariance. In this work we show that the requirement for the simultaneous fulfillment of mutual locality of the left-moving  vertices of physical states with the space-time symmetry generators and of right-moving vertices with generators of $E(8)\times E(6)$-gauge symmetry, which arises after some special reduction together with the requirement of mutual locality of complete (left-right) vertices of physical states among themselves leads to the same Gepner models.}


\begin{document}
\maketitle

\section{Introduction}
\label{sec:intro}

Superstring \cite{GS} and Heterotic string \cite{GHMR}, \cite{CHSW} 
 approaches to constructing a $d=10$ theory of quantum gravity 
and $d=10$ Grand Unified Theory are distinguished by their combination of uniqueness and mathematical beauty.

The discovery by Gepner \cite{Gep}, see also \cite {BLT}, of the equivalence of the compactification of $6$ dimensions of $10$-dimensional spacetime on Calabi-Yau manifolds and the compactification on $N=2$ SCFT with total central charge $c=9$   leads to the uniqueness and exact solvability of the constructed $d=4$ string models that have space-time supersymmetry and needed gauge symmetry.

Heterotic 4-dimensional Gepner string models result from the coupling of a left-handed fermion string derived from $N=1$ CFT, the additional 6 dimensions of which are compactified on $M_{\vec{k}}$ product of  minimal $N=2$ SCFT models with total central charge $9$ and a right-moving bosonic string whose additional 9 dimensions are also compactified on $ M_{\vec{k}}$ and whose remaining 13 dimensions form a torus $E(8)\times SO(10)$ algebra.

Such Heterotic string models have $N=1$ Space-time symmetry arising after the GSO reduction  in its left-moving part and $E(8)\times E(6)$ Gauge symmetry arising after a similar reduction of its right-hand side.
It is known that these symmetries are necessary for phenomenological reasons,
they are used to create Grand Unified theories.

We show that the requirement for the simultaneous fulfillment of mutual locality of the left-moving  vertices  with the space-time symmetry generators and of right-moving vertices with generators of $E(8)\times E(6)$-gauge symmetry
together with 
the requirement of mutual locality of complete (left-right) vertices of physical states among themselves and other demands of the Conformal bootstrap \cite{BPZ} lead to Gepner models obtained from the requirement of simultaneous fulfillment of space-time symmetry with modular invariance.

Based on these requirements, we explicitly construct the  physical states of the theory in three steps. 

At the first step,  in the left sector  we find  the generators of 
$N=1$ space-time supersymmetry and the set of left-moving  physical states that are mutually local with them.

At the second  step, in the  right sector we find the generators of $E(6)$ symmetry  and the set of right-moving physical states that are mutually local with them.

At the third step, we find the products of the left- and right-moving vertices of the states  obtained in this way, which are mutually local to each other.

We generalize the construction to include compactifications on orbifolds. At the end of the article we provide expressions for the states of massless matter for the quintic to demonstrate our approach.

\section{Left-moving sector, N=1  SCFT}
\label{sec:2}

In the left sector we have the product of  4-dimensional $N=1$ CFT for a space-time subsector consisting of $4$ bosons and $4$ Majorano fermions  with the central charge of 6 and  $N=1$ CFT for the compact subsector with the central charge of 9, so that the total central charge in the left sector is $c =15$.

Thus $N=1$ CFT of space-time factor is a theory of 4 free bosons $X^{\mu}(z)$ and 4 Majorana fermions $\psi^{\mu}(z)$
\be
\begin{aligned}
	&X^{\mu}(z)X^{\nu}(0)=-\eta^{\mu\nu}\log{z}+..., 
	\nmb
	&\psi^{\mu}(z)\psi^{\nu}(0)=\eta^{\mu\nu}z^{-1}+...,
	\label{2.XpsiOPE}
\end{aligned}
\en

$N=1$ CFT for the  compact factor  realized as a product $N=2$ minimal models \cite { FZ} with the total central charge $9$\footnote{Here and below we consider the product of five minimal models for simplicity.}	
\begin{equation}
	\begin{aligned}
		&M_{\vec{k}}=\prod_{i=1}^{5}M_{k_{i}}, \ c_{i}=\frac{3k_{i}}{k_{i}+2}, \ \sum_{i}c_{i}=9.
	\end{aligned}
\end{equation}

Recall that $N=2$ superconformal minimal model contains the finite number of primary fields \cite{FZ}
\begin{equation}
	\begin{aligned}
		\Phi^{l}_{q,s}(z), \ l=0,...,k, \ l+q+s=0\,\, \textrm{mod} \ 2, \ s=0,1,2,3,
		\\
		\Delta=\frac{l(l+2)-q^{2}}{4(k+2)}+\frac{s^{2}}{8}, \
		Q=\frac{q}{k+2}-\frac{s}{2},
	\end{aligned}
\end{equation}
where in NS sector, $s=0,2$ and $s=1,3$ for the fields from R sector.

The total left-moving $N=1$ Virasoro superalgebra of the matter is the diagonal subalgebra in the direct sum of the $N=1$ Virasoro algebras of space-time degrees of freedom and the minimal models of compact factor
\begin{equation}
	\begin{aligned}
		&T_{mat}(z)=T_{st}(z)+T_{int}(z), \ G_{mat}(z)=G_{st}(z)+G_{int}(z), 
		\\
		&T_{st}=-\frac{1}{2}\d X^{\mu}(z) \d X_{\mu}(z)-\frac{1}{2}\psi^{\mu}(z)\d\psi_{\mu}(z), 
		\\
		&G_{st}(z)=\d X^{\mu}\psi_{\mu}(z),
		\\
		&T_{int}=\sum_{i=1}^{5}T_{i}(z),
		\
		G_{int}(z)=\sum_{i=1}^{5}(G^{+}_{i}+G^{-}_{i})(z).
		\label{LVir}
	\end{aligned}		
\end{equation}
The currents $T_{i}(z)$, $G^{\pm}_{i}(z)$ together with the $U(1)$ current $J_{i}(z)$ form $N=2$ Virasoro superalgebra of the minimal model $M_{i}$.

The $N=1$ Virasoro superalgebra action is correctly defined on the product of only $NS$-representations or on the product of only $R$-representations.

We use BRST approach to define the physical states. The BRST charge is given by the integral 
\begin{equation}
	Q_{\textrm{BRST}}=\oint dz (cT_{mat}+\gamma G_{mat}+\fr12(cT_{gh}+\gamma G_{gh})),
	\label{BRST}
\end{equation}
where we introduced the  ghost fields and $N=1$ Virasoro superalgebra of the ghosts
\begin{equation}
	\beta(z)\gamma(0)=-z^{-1}+..., \ \ b(z)c(0)=z^{-1}+....
	\label{ghope}
\end{equation}
\begin{equation}
	\begin{aligned}
		&T_{\textrm{gh}}=-\d bc-2b\d c-\frac{1}{2}\d\beta\gamma-\frac{3}{2}\beta\d\gamma,
		\\
		&G_{\textrm{gh}}=\d\beta c+\frac{3}{2}\beta\d c-2 b\gamma.
		\label{1.Virgh}
	\end{aligned}
\end{equation}
The ghost  space of states is characterized by the vacuum $V_{q}(z)$, which can be realized as a free scalar field exponent
\begin{equation}
	\begin{aligned}
		&V_{q}(z)=\exp{(q\phi(z))},\
		\phi(z)\phi(0)=-\log(z)+...,
		\\
		&\beta(z)V_{q}(0)\sim O(z^{q}),\
		\gamma(z)V_{q}(0)\sim O(z^{-q}).
		\label{picvac}
	\end{aligned}
\end{equation}

The left-moving vertex can be written as
\begin{equation}
	\begin{aligned}
		&V^{\vec{l}}_{\vec{\mu}_{L}}=
		P_{gh}(\beta,\gamma,b,c)P_{st}(\d X^{\mu},\d H_{a}) P_{int}
		\exp{[q\phi+ \im\lm^{a}H_{a}+ \im p_{\mu}X^{\mu}(z)]}
	{\Phi}^{\vec{l}}_{\vec{q}_{L},\vec{s}_{L}}(z),
	\\
	&\vec{\mu}_{L}=(q,\vec{\lm},\vec{q}_{L},\vec{s}_{L}), \ \vec{q}_{L}=(q^{1}_{L},...,q^{5}_{L}), \ \vec{s}_{L}=(s^{1}_{L},...,s^{5}_{L}), 
	\\ 
	&Q_{L}^{i}:=\frac{q_{L}^{i}}{k_{i}+2}-\frac{s_{L}^{i}}{2} 
	+ \text{even},
	\
	s^{i}_{L}=0,1,2,3, \ l_{i}+q^{i}_{L}+s^{i}_{L}=0 \mod 2. 
	\label{2.LVert}
\end{aligned}
\end{equation}
$P_{gh}, P_{st}, P_{int}$ are the polynomials of the corresponding fields and their derivatives and the polynomial $P_{int}$ contains only even number of fermionic fields.

Here we have bosonized fermions:
\begin{equation}
\begin{aligned}
	&H_{a}(z)H_{b}(0)=-\delta_{ab}\log{(z)}+..., \ a,b=1,2.
	\\
	&\frac{1}{\sqrt{2}}(\pm\psi^{0}+\psi^{1})=\exp{[\pm\im H_{1}]},\
	\frac{1}{\sqrt{2}}(\psi^{2}\pm\im\psi^{3})=\exp{[\pm\im H_{2}]}.		
	\label{ferm}
\end{aligned}
\end{equation}
Thus, the exponential $\exp{[\im\lm^{a}H_{a}]}$ corresponds to the fully filled space-time fermionic state which is characterized by the $SO(1,3)$ weight $\vec{\lm}$. 

Using the  definition
\begin{equation}
	\vec{\mu}_{L}\cdot\vec{\mu}_{L}':=-qq'+\vec{\lm}\cdot\vec{\lm}'
	-\fr12\sum_{i} ( \frac{q_i q'_i}{k_{i}+2}-\frac{s_i s'_i}{2})
\end{equation}	
the corresponding restrictions on the vertices (\ref{2.LVert}) that ensure the action of the diagonal $N=1$ Virasoro superalgebra (\ref{LVir}) can be rewritten in the form
$\vec{\beta}_{j}\cdot\vec{\mu}_{L}\in\mathbb{Z}$, 
\begin{equation}
	\begin{aligned}
		&  \ \vec{\beta}_{1}:=(1;1,0;0,...,0;0,...,0),
		\\
		& \ \vec{\beta}_{2}:=(1;0,1;0,...,0;0,...,0),
		\\
		&  \ \vec{\beta}_{3}:=(1;0,0;0,...,0;2,0,0,0,0),\\
		&\dots\dots\dots\dots\dots\dots\dots\dots\dots\dots,\\
		& \ \vec{\beta}_{7}:=(1;0,0;0,...,0;0,0,0,0,2),	
	\end{aligned}
\end{equation}
where  the dimensions of the vectors $\vec{\beta}_j$ is $13$.
 
As we will see below, it is important that the following equations also hold
\begin{equation}
	\begin{aligned}
		&\vec{\beta}_{j}\cdot \vec{\beta}_{0}\in\mathbb{Z}, \ j=1,...,7,
		\\
		&\vec{\beta}_{0}:=(\frac{1}{2};\frac{1}{2},\frac{1}{2};1,...,1;1,...,1).
	\end{aligned}
\end{equation}

\section{Massless physical left movers and space-time supersymmetry}\label{sec:3}
Imposing the BRST-invariance requirement, we find the following  massless left vertices in the sector NS with the canonical picture number $(-1)$ and in the sector R with the canonical  picture number $(-\fr12)$.
We get the left-moving  vertex  of massless vector boson 
in the $(-1)$ picture
\begin{equation}
	\exp(-\phi(z)) \xi_{\mu}(p)\psi^{\mu}(z) 
	\exp\left(ip_{\mu}X^{\mu}(z)\right), 
	\quad
	\xi_{\mu}(p)p^{\mu}=0.
	\label{3.Vect}
\end{equation}

We also find the left-moving vertex of massless scalar bosons
\begin{equation}
	\begin{aligned}
		&V^{c}_{\vec{l},p}=\exp{(-\phi)}\Phi^{\vec{l}}_{\vec{l},\vec{0}}\exp{(\im p_{\mu}X^{\mu}(z))}, \ \sum_{i=1}^{5}\frac{l_{i}}{k_{i}+2}=1,
		\\
		&V^{a}_{\vec{l},p}=\exp{(-\phi(z))}\Phi^{\vec{l}}_{-\vec{l},\vec{0}}(z)\exp{(\im p_{\mu}X^{\mu}(z))},
		\ \sum_{i=1}^{5}\frac{l_{i}}{k_{i}+2}=1,
	\end{aligned}
\end{equation}
where $\Phi^{\vec{l}}_{\vec{l},\vec{0}}(z)$, $\Phi^{\vec{l}}_{-\vec{l},\vec{0}}(z)$ are chiral and anti-chiral primary states from the compact factor.

In the picture $(-\fr12)$ we find the vertices of massless spinors
\begin{equation}
	\begin{aligned}
		&S^{\pm}_{\xi,p}=\sum_{\vec{\sigma}}\xi_{\vec{\sigma}}(p)\exp(-\fr12\phi + \im {\vec{\sigma}}\cdot \vec{H}
		\pm\frac{\im}{2}\sum_{i}\frac{k_{i}\phi_{i}}{\sqrt{k_{i}(k_{i}+2)}}) \exp{(\im p_{\mu}X^{\mu}(z))},
		\\ 
		&\sigma^{a}=\pm\fr12,  \ \sum_{a=1}^{2}\sigma^{a}=\pm 1, \ p^{2}=0,
		\\
		&\dot{S}^{\pm}_{\xi,p}=\sum_{\dot{\vec{\sigma}}}\xi_{\dot{\vec{\sigma}}}(p)\exp(-\fr12\phi(z) + \im \dot{\vec{\sigma}}\cdot \vec{H} \pm\frac{\im}{2}\sum_{i}\frac{k_{i}\phi_{i}(z)}{\sqrt{k_{i}(k_{i}+2)}}) \exp{(\im p_{\mu}X^{\mu}(z))},\\
		&\dot{\sigma}^{a}=\pm\fr12,  \ \sum_{a=1}^{2}\dot{\sigma}^{a}=0, \ p^{2}=0,
	\end{aligned}
\end{equation}
where fields $\phi_i$ appear as a result of bosonization of $U(1)$ currents in minimal models $M_{k_i}$.

By setting $p=0$ at these vertices, we obtain currents whose integrals are the supergenerators of the space-time supersymmetry
\begin{equation}
	\begin{aligned}
		&S^{\pm}_{\sigma}(z)=\exp(-\fr12\phi(z) + \im {\sigma}^{a}H_{a}(z) 
		\pm\frac{\im}{2}\sum_{i}\frac{k_{i}\phi_{i}(z)}{\sqrt{k_{i}(k_{i}+2)}}),
		\sigma^{a}=\pm\fr12,  \ \sum_{a=1}^{2}\sigma^{a}=\pm 1,\\
		&S^{\pm}_{\dot{\sigma}}(z)=\exp(-\fr12\phi(z) + \im \dot{{\sigma}}^{a}H_{a}(z)  \pm\frac{\im}{2}\sum_{i}\frac{k_{i}\phi_{i}(z)}{\sqrt{k_{i}(k_{i}+2)}}), 
		\dot{ \sigma}^{a}=\pm\fr12,  \ \sum_{a=1}^{2}\dot{\sigma}^{a}=0.
	\end{aligned}
\end{equation}
These currents are nothing else but the product of the Ramond vacua of the three sub-theories of the string model under consideration. In particular, the exponential factors of the internal model $M_{\bk}$ are products $\prod_{i}U_{i}^{\fr12}$ of the spectral flow  operators \cite {SS} for each minimal model $M_{k_{i}}$.

It is easy to verify that each of the four currents 
$S^{+}_{\sigma}$, $S^{+}_{\sigma}$,  $S^{+}_{\dot{\sigma}}$ and $S^{-}_{\dot{\sigma}}$,  belongs to the two-dimensional representation 
of the $SO(1,3)$ Lie algebra of space-time symmetry, namely, each of them  is  a  Weyl spinor.
Two currents $S^{+}_{\sigma}$ and $S^{-}_{\sigma}$ have the same chirality, say left-handed,  and the other two, $S^{+}_{\dot{\sigma}}$ and $S^{-}_{\dot{\sigma}}$, have opposite chirality, respectively right-handed.
For our construction, the mutual locality of these currents is important.
A simple check shows that the current  $S^{+}_{\sigma}$ is mutually local with $S^{-}_{\dot{\sigma}}$  and  $S^{-}_{\sigma}$ is mutually 
local with $S^{+}_{\dot{\sigma}}$. 

The four supercharges, defined by the integration of these  currents, define, together with generators of  Poincaré algebra operators, two $N=1$ super-Poincaré algebras of opposite chiralities. The currents $S^{+}_{\sigma}$ and $S^{-}_{\dot{\sigma}}$ define
super generators  of  a $N=1$ super Poincaré for the case 
of left chirality as
\begin{equation}
	\begin{aligned}
		&{\cal{Q}}^{+}_{\sigma}= \oint S^{+}_{\sigma}  dz 
		=\oint dz \exp{[-\fr12\phi+\im {\sigma}^{a}H_{a}+\frac{\im}{2}
			\sum_{i}\frac{k_{i}\phi_{i}}{\sqrt{k_{i}(k_{i}+2)}}]},\\ 
		&{\cal{Q}}^{-}_{\dot{\sigma}}= \oint S^{-}_{\dot{\sigma}} dz
		=\oint dz \exp{[-\fr12\phi+\im \dot{\sigma}^{a}H_{a}-\frac{\im}{2}
			\sum_{i}\frac{k_{i}\phi_{i}}{\sqrt{k_{i}(k_{i}+2)}}]}. 
	\end{aligned}
	\label{3.LSUSYa}
\end{equation}

Instead of the mutually local pair $S^{+}_{\sigma}$ and $S^{-}_{\dot{\sigma}}$ we can take another pair $S^{-}_{\sigma}$ and $S^{+}_{\dot{\sigma}}$. As a result of either of these two different options, we obtain  $d=4$ $N=1$ spacetime supersymmetry. For each of these two options, we have to leave from the BRST-invariant fields only those fields that are mutually local with respect to the chosen supercharges.

It is easy to see that the SUSY currents $S^{+}_{\sigma}, S^{-}_{\dot{\sigma}}$  are mutually local with the vertex $ V^{\vec{l}}_{\vec{\mu}_{L}} $ if
\begin{equation}
	\fr12 q+\sum_{a}\sigma^{a}\lm^{a}+\fr12\sum_{i=1}^{5}
	(\frac{q^{i}_{L}}{k_{i}+2}-\fr12s^{i}_{L})\in \mathbb{Z}.
	\label{3.GSO}
\end{equation}
\begin{equation}
	\fr12 q+\sum_{a}\sigma^{a}\lm^{a}+\fr12\sum_{i=1}^{5}Q_{L}^{i}
	\in \mathbb{Z}
\end{equation}
where $Q_{L}^{i}$ the $U(1)$ charge in the model $M_{k_{i}}$.

This equation can be rewritten as
\begin{equation}
		\vec{\beta}_{0}\cdot\vec{\mu}_{L}\in\mathbb{Z},\ 
		\text{where} \ \vec{\beta}_{0}:=(\fr12;\fr12,\fr12;1,...,1;1,...,1).
		\label{3.GSO1}
\end{equation}
It  is equivalent to  the GSO condition.

From the GSO equation and the requirement of compatibility with  $N=1$ Virasoro  it follows that the total internal charges $Q_{L}:=\sum_{i}Q^{i}_{L}$ of the selected vertices are integers or half-integers.

One can get more explicit structure of GSO projected vertices if one takes into account that the weights $\vec{\lm}$ of the algebra $SO(2n)$ split into the four conjugacy classes
\begin{equation}
	\begin{aligned}
		& (0): (0,0,0,...,0) + \text{any root};
		\\
		&  (V): (1,0,0,...,0)+ \text{any root};
		\\
		& (S):  (\frac{1}{2},\frac{1}{2},\frac{1}{2},...,\frac{1}{2})+
		\text{any root};
		\\
		& (C):  (-\frac{1}{2},\frac{1}{2},\frac{1}{2},...,\frac{1}{2})+
		\text{any root}.
		\label{3.class}
	\end{aligned}
\end{equation}
From the requirement of compatibility between the action of $ N=1$  diagonal Virasoro algebra and  GSO equation  follows that in NS sector the weights $\vec{\lm}$ of the vertex (\ref{2.LVert}) fall into classes $[0]$ and $[V]$ and
in R sector the weights $\vec{\lm}$ fall into $[S]$  and $[C]$ of $SO(1,3)$.

As a result, we obtain the following agreement between pictures $q$, conjugacy classes $\vec{\lm}$ and sums of $U( 1)$ charges in a compact sector for 4 classes of admissible left vertices
\begin{equation}
	\begin{aligned}
		& \sum_{i}Q^{i}_{L}\in 2\mathbb{Z}+1  
		\Rightarrow q=-1,
		\sum_{a}\sigma^{a}\lm^{a}=0 \mod \  \mathbb{Z}, 
		(\vec{\lm}\in[0]),
		\\
		& \sum_{i}Q^{i}_{L}\in 2\mathbb{Z}  
		\Rightarrow q=-1,
		\sum_{a}\sigma^{a}\lm^{a}=\fr12 \mod \  \mathbb{Z}, 
		(\vec{\lm}\in[V]),
		\\ 
		& \sum_{i}Q^{i}_{L}\in 2\mathbb{Z}-\fr12 
		\Rightarrow q=-\fr12,
		\sum_{a}\sigma^{a}\lm^{a}=\fr12 \mod \  \mathbb{Z}, 
		(\vec{\lm}\in[S]),
		\\ 
		& \sum_{i}Q^{i}_{L}\in 2\mathbb{Z}+\fr12 
		\Rightarrow q=-\fr12,
		\sum_{a}\sigma^{a}\lm^{a}=0 \mod \  \mathbb{Z}, 
		(\vec{\lm}\in[C]).
	\end{aligned}
	\label{3.LGSOsol}
\end{equation}

If $\vec{\mu}_L$ obeys the GSO equations, then the new vector $\vec{\mu}_L+\delta\vec{\mu}_L$ will also satisfy the GSO equations if
\begin{equation}
	\vec{\beta}_{0}\cdot\delta\vec{\mu}_{L}\in\mathbb{Z}.
\end{equation}

In particular, the deformations generated by $\vec{\beta}_{0}$,...,$\vec{\beta}_{7}$ preserve the GSO equation by simply changing the descendant states accordingly.

The additional possible deformations that preserve the GSO equation are associated with the vectors $w_{i}\in\mathbb{Z}_{k_{i}+2}$, which satisfy the equation
\begin{equation}
	\sum_{i=1}^{5}\frac{w_{i}}{k_{i}+2}\in\mathbb{Z},
	\label{3.Accept}
\end{equation}
if such solutions other than $\vec{w}_0:=(1,1,1,1,1)$ exist.
We will use these deformations below  in orbifold compactification.

\section{Right-moving sector, N=0 CFT}
\label{sec:4}
Space-time factor in the right-moving sector with $c=4$ contains only bosonic fields $\bar{X}^{\mu}(\bar{z})$. 

In order to get critical string in the right-moving sector we add the bosonic fields $Y_{I}(\bar{z})$, $I=1,...,8$ which are compactified on  the torus of the algebra $E(8)$ with $c=8$ and the bosons $\bar{H}_{\al}(\bar{z})$, compactified on the torus of the algebra $SO(10)$ with $c=5$.

The remaining contribution to the critical dimension is given by the right-moving part of the compact factor $M_{\vec{k}}$.

Right-moving energy-momentum tensor is 
\begin{equation}
	\bar{T}_{mat}(\bar{z})=\fr12(\eta_{\mu\nu}\bar{\d}\bar{X}^{\mu}\bar{\d}\bar{X}^{\nu}+(\bar{\d}Y_{I})^{2}+(\bar{\d}\bar{H}_{\al})^{2})+\bar{T}_{int}(\bar{z}).
\end{equation}
We use the BRST approach introducing right-moving ghosts:
\begin{equation}
	\bar{b}(\bar{z})\bar{c}(0)=\bar{z}^{-1}+...
\end{equation}
and BRST operator
\begin{equation}
	\begin{aligned}
		&\bar{Q}_{BRST}=\oint d\bar{z} \bar{c}(\bar{T}_{mat}+\fr12\bar{T}_{gh}) , 
		\\
		&\bar{T}_{gh}=\bar{c}\bar{\d} \bar{b}-2\bar{b}\bar{\d}\bar{c}.
		\label{4.RBRST}
	\end{aligned}
\end{equation}

The general right-moving vertex can be written as
\begin{equation}
	\begin{aligned}
		&\bar{V}^{\vec{l}}_{\vec{\mu}_{R}}(\bar{z})=P_{gh}(\bar{b},\bar{c})P_{st}(\bar{\d} \bar{X}^{\mu})P_{int}(\bar{\d}\bar{Y}^{I},\bar{\d}\bar{H}^{\al},\bar{T}_{i}, \bar{J}_{i}, \bar{G}^{\pm}_{i})
		\\
		&\exp{[\im \eta_{I}\bar{Y}^{I}+
			\im\Lm_{\al}\bar{H}^{\al} + \im p_{\mu}\bar{X}^{\mu}(z)]}
		\bar{\Phi}^{\vec{l}}_{\vec{q}_{R},\vec{s}_{R}}(\bar{z}),
		\\
		&\vec{\mu}_{R}=(\vec{\eta},\vec{\Lm},\vec{q}_{R},\vec{s}_{R}), \ \vec{q}_{R}=(q^{1}_{R},...,q^{5}_{R}), \ \vec{s}_{R}=(s^{1}_{R},...,s^{5}_{R}),
		\label{4.RVert}
	\end{aligned}
\end{equation}
where $\vec{\eta}$ is a vector of the $E(8)$ root  lattice, and $\Lm$ is the vector of the $SO(10)$ weight lattice.

Among the vertices (\ref{4.RVert}) we find BRST-invariant massless ones. First of all this is $SO(1,3)$ vector $V^{\mu}(\bar{z})=\im\bar{\d}\bar{X}^{\mu}(\bar{z})$,
Then we find the currents of $E(8)$ algebra
\begin{equation}
	V^{I}(\bar{z})=\im\bar{\d}\bar{Y}^{I}(\bar{z}), \ I=1,...,8,
	\
	V_{\vec{\eps}}(\bar{z})=\exp{[\im \eps_{I}\bar{Y}^{I}]}(\bar{z}), \ \vec{\eps}^{2}=2,
	\label{4.E8}
\end{equation}
\begin{equation}
	\begin{aligned}
		\vec{\eps}=
		\begin{cases}(\pm 1,\pm 1,0,0,0,0,0,0)+ \text{permutations},
			\\
			(\pm\frac{1}{2},...,\pm\frac{1}{2})+ \text{permutations},\, 
			\text{even  number  of}  +\frac{1}{2},
		\end{cases}
	\end{aligned}
\end{equation}
where the vectors $\vec{\eps}$ are the roots of $E(8)$ algebra.

Also there are the currents of $SO(10)$  algebra:
\begin{equation}
	\begin{aligned}
		&V^{\al}(\bar{z})=\im\bar{\d}\bar{H}^{\al}(\bar{z}), \ \al=1,...,5,
		\\
		&V_{\vec{\rho}}(\bar{z})=\exp{[\im \rho_{\al}\bar{H}^{\al}]}(\bar{z}), \ \rho_{\al}=\pm 1, \ \sum (\rho_{\al})^{2}=2,
		\label{4.SO10}
	\end{aligned}
\end{equation}
where the vectors $\vec{\rho}$ 
\begin{equation}
	\vec{\rho}=(\pm 1, \pm 1,0,0,0)+ \text{permutations}
\end{equation}
are the roots of $SO(10)$.

As well as the currents of  $U(1)^{4}$ algebra
\begin{equation}
	I_{j}(\bar{z})=\im\sqrt{\frac{k_{j}}{k_{j}+2}}\bar{\d}\bar{\phi}_{j}(\bar{z})-\frac{k_{j}}{3(k_{j}+2)}J_{int}(\bar{z}), \ j=1,...,4.
	\label{4.EM}
\end{equation}

In the right-moving sector there are also massless $SO(10)$ spinors
\begin{equation}
	\begin{aligned}
		&\Sigma^{\pm}_{\om}(\bar{z})=\exp{[\im \om_{\al}\bar{H}^{\al}]}\exp{[\pm\frac{\im}{2}\sum_{i}\frac{k_{i}\bar{\phi_{i}}}{\sqrt{k_{i}(k_{i}+2)}}]}, 
		\\
		&\om_{\al}=\pm\fr12, \ \sum \om_{\al}=\fr12 \mod \ 2\mathbb{Z},
		\\
		&\Sigma^{\pm}_{\dot{\om}}(\bar{z})=\exp{[\im \dot{\om}_{\al}\bar{H}^{\al}]}\exp{[\pm\frac{\im}{2}\sum_{i}\frac{k_{i}\bar{\phi_{i}}}{\sqrt{k_{i}(k_{i}+2)}}]}, 
		\\
		&\dot{\om}_{\al}=\pm\fr12, \ \sum \dot{\om}_{\al}=-\fr12 \mod \ 2\mathbb{Z}.
	\end{aligned}
\end{equation}
$\Sigma^{+}_{\om}$ is mutually local with $\Sigma^{-}_{\dot{\om}}$. And $\Sigma^{-}_{\om}$ is mutually local 
with $\Sigma^{+}_{\dot{\om}}$.

As it was found by Gepner \cite{Gep}, the 45 currents of $SO(10)$ algebra  together with the $32$ spinor currents $\Sigma^{+}_{\om}(\bar{z})$, $\Sigma^{-}_{\dot{\om}}$ and the $U(1)$ current
\begin{equation}
	\bar{J}_{int}=\im\sum_{i}\sqrt{\frac{k_{i}}{k_{i}+2}}\bar{\d}\bar{\phi}_{i}(\bar{z})
\end{equation}
form $E(6)$ current algebra at level 1.

One can rewrite the $E_{6}$ currents in terms of simple roots of
 $E_{6}$
\begin{equation}
	\begin{aligned}
		&\vec{\al}_{i}=e_{1}-e_{2},...,\vec{\al}_{4}=e_{4}-e_{5}, \vec{\al}_{5}=e_{4}+e_{5},
		\\
		&\vec{\al}_{6}=-\frac{1}{2}(e_{1}+...+e_{5})+\frac{\sqrt{3}}{2}e_{6},
	\end{aligned}
\end{equation}
where $e_{i}$ are the orthonormal basic vectors in  $\mathbb{R}^{6}$. So that the  Cartan subalgebra currents are
\begin{equation}
	\begin{aligned}
		&h_{j}(\bar{z})=\im\vec{\al}_{j}\cdot\bar{\d}\vec{\bar{H}}(\bar{z}), \ j=1,...,6, \text{where}
		\\
		&\vec{\bar{H}}(\bar{z})=(\bar{H}_{1}(\bar{z}),...,\bar{H}_{5}(\bar{z}), \bar{H}_{6}(\bar{z})), 
		\\
		& \bar{H}_{6}(\bar{z})=\sum_{i}\sqrt{\frac{k_{i}}{3(k_{i}+2)}}
		\bar{\d}\bar{\phi}_{i}(\bar{z}).
		\label{4.E6C}
	\end{aligned}
\end{equation}
The currents of ladder $E(6)$ operators are given by
\begin{equation}
	\begin{aligned}
		& E_{j}(\bar{z})=\exp{[\im\vec{\al}_{j}\vec{\bar{H}}]}(\bar{z}),
		\\
		& F_{j}(\bar{z})=\exp{[-\im\vec{\al}_{j}\vec{\bar{H}}]}(\bar{z}),
		\ j=1,...,6.
		\label{4.E6r}
	\end{aligned}
\end{equation}
The other $E(6)$ currents can be generated from (\ref{4.E6C}), (\ref{4.E6r}) by the OPE's.

$E(6)$ is known to be considered  as a possible Grand Unified gauge group, which after spontaneous breaking gives rise to the $SU(3)\times SU(2)\times U(1)$ gauge group of the Standard Model.

Taking this into account, we would like to emphasize that the $E(6)$-algebra appears in the right-hand sector in some sense inevitably after we introduce $N=1$ SUSY into the left-hand sector.

\section{$E(6)$  symmetry and right-moving $"GSO"$ equations}
\label{sec:5}
To obtain a set of states compatible with the action
of $E(6)$ we need to select vertices that are not only mutually local with the currents $SO(10)$, but also with additional currents $E(6)$ (\ref{4.E6r}). It requires in particular the mutual locality of the currents $\Sigma^{+}_{\om}(\bar{z})$, $\Sigma^{-}_{\dot{\om}}(\bar{z})$ with vertex (\ref{4.RVert}). These requirements are satisfied if the following consistency equations ("GSO" equations) are satisfied.
\begin{equation}
	\om\cdot\Lm+ \fr12\sum_{i}
	(\frac{q^{i}_{R}}{k_{i}+2}-\fr12 s^{i}_{R})
	\in \mathbb{Z}.
	\label{5.GSO}
\end{equation}

These "GSO" equations mean  that $6$-vector
$(\Lm,\frac{1}{\sqrt{3}}\sum_{i=1}^{5}Q^{i}_{R})$, where 
$Q^{i}_{R}=\frac{q^{i}_{R}}{k_{i}+2}-\fr12 s^{i}_{R}$ 
is the $U(1)$ charge   of the state in the $i$- factor must be an $E(6)$ weight lattice vector.

From the "GSO" equations in the right-moving sector we  get that the $SO(10)$ parts of the solutions of the right vertices, that fall into one of the four conjugacy classes, determine the sixth, internal component as follows
\begin{equation}
	\begin{aligned}
		&\vec{\Lm}\in[0]\Rightarrow \sum_{\al}\om^{\al}\Lm^{\al}=0 \mod \  1 \Rightarrow \sum_{i}Q^{i}_{R}\in 2\mathbb{Z}, 
		\\
		&\vec{\Lm}\in[V]\Rightarrow \sum_{\al}\om^{\al}\Lm^{\al}=\fr12
		\mod \ 1 \Rightarrow \sum_{i}Q^{i}_{R}\in 2\mathbb{Z}+1,
		\\
		&\vec{\Lm}\in[S]\Rightarrow \sum_{\al}\om^{\al}\Lm^{\al}=\frac{1}{4} \mod \  1 \Rightarrow \sum_{i}Q^{i}_{R}\in 2\mathbb{Z}-\fr12,
		\\		
		&\vec{\Lm}\in[C]\Rightarrow \sum_{\al}\om^{\al}\Lm^{\al}=-\frac{1}{4} \mod \ 1 \Rightarrow \sum_{i}Q^{i}_{L}\in 2\mathbb{Z}+\fr12.
	\end{aligned}
	\label{5.RGSOsol}
\end{equation}

\section{Mutual locality of the full physical  vertices}
\label{sec:6}
Our rule for “gluing” left vertices with right ones is the requirement of mutual locality of full vertices. So the physical vertices of the Heterotic string must  be BRST invariant from the left and from the right, obey the left-moving $GSO$ equation (\ref{3.GSO1}) and the right-moving $"GSO"$ equation (\ref{5.GSO}) and be mutually local to each other.

We start our search of full mutually local vertices among the so called “quasi-diagonal”. The “quasi-diagonal” full vertices are given by the product of GSO-invariant left-moving and “GSO”-invariant right-moving factors
\begin{equation}
	\begin{aligned}
		&{\cal{V}}^{\vec{l}}_{\vec{\mu}_{L},\vec{\mu}_{R}}(z,\bar{z})=
		V^{\vec{l}}_{\vec{\mu}_{L}}(z)\times\bar{V}^{\vec{l}}_{\vec{\mu}_{R}}(\bar{z})=
		\\
		&P^L_{gh}(\beta,\gamma,b,c)P^L_{st}(\d X^{\mu},\d H_{\al})
		 P^L_{int}(\dots)
		\exp{[q\phi+\im\lm^{a}H_{\al}]}
		\Phi^{\vec{l}}_{\vec{q}_{L},\vec{s}_{L}}(z)\times
		\\
		&\times P^R_{gh}(\bar{b},\bar{c})P^R_{st}(\bar{\d} \bar{X}^{\mu})
		  P^R_{int}(\dots)
		\exp{[\im \epsilon^{I}\bar{Y}_{I}+\im\Lm^{a}\bar{H}_{a}]}
		\bar{\Phi}^{\vec{l}}_{\vec{q}_{R}, \vec{s}_{R}}(\bar{z}),
		\label{6.DiagV}
	\end{aligned}
\end{equation}
where 
\be
\vec{l}_{L}=\vec{l}_{R}, \vec{q}_{L}=\vec{q}_{R}, \vec{s}_{L}=\vec{s}_{R}.
\en
The product of two such vertices after moving one around the other receives a complex factor, the  phase of which has the following form
\be
2\pi \im (\vec{\mu}_{L}\cdot\vec{\mu}_{L}'-
\vec{\mu}_{R}\cdot\vec{\mu}_{R}')=
2\pi \im (-qq'+\vec{\lm}\cdot\vec{\lm}'-\vec{\Lm}\cdot\vec{\Lm}').
\en

The mutually locality requirement
\be
\vec{\mu}_{L}\cdot\vec{\mu}_{L}'- \vec{\mu}_{R}\cdot\vec{\mu}_{R}'\in \mathbb{Z}
\en
imposes certain correlations between  the classes  $\vec{\lm}$  and $\vec{\Lm}$. The reason for this is that  GSO equations and the requirement of compatibility with the $N=1$ Virasoro action formulated above lead to a correlation between the picture number $q$, the conjugacy classes  of $\vec{\lm}$ and the total internal charges in the left sector.

The same is correct  for the correlation between  classes of  $\vec{\Lm}$ and total internal charges in the right-moving sector because of  the right "GSO" equations.

Considering also that for quasi-diagonal full vertices $\sum_{i}Q^{i}_L =\sum_{i}Q^{i}_R$, we get four types of them, which meet the following requirements
\begin{equation} 
	\begin{aligned}
		&\sum_{i}Q^{i}\in 2\mathbb{Z}\Rightarrow q=-1, \ \lm\in[V], \ \Lm\in[0],
		\\	
		&\sum_{i}Q^{i}\in 2\mathbb{Z}+1\Rightarrow q=-1, \ \lm\in[0], \ \Lm\in[V],
		\\	
		&\sum_{i}Q^{i}\in 2\mathbb{Z}+\frac{1}{2}\Rightarrow q=-\frac{1}{2}, \ \lm\in[C], \ \Lm\in[C],
		\\
		&\sum_{i}Q^{i}\in 2\mathbb{Z}-\frac{1}{2}\Rightarrow q=\frac{1}{2}, \ \lm\in[S], \ \Lm\in[S].
		\label{6.Modinv}
	\end{aligned}
\end{equation}

These vertices are mutually local due to the correlation  of the internal charges $\sum Q^{i}$, $SO(1,3)$ weights $\lm$ and $SO(10)$ weights $\Lm$\footnote{Note that here we observe the replacement of the singlet and vector $SO(1,3)$ for left movers with the vector and singlet $SO(10)$ for right movers, which is necessary to ensure modular invariance of the theory, as it was shown by Gepner.}.

However, the set of the full mutually “quasi-diagonal” vertices obtained in this way has one drawback. It does not satisfy the requirement of space-time supersymmetry.

To fix this problem, we will use the following argument. The superpartners are generated by   $\vec{\beta}_{0}$ shifts $\vec{\mu}_{L}\rightarrow \vec{\mu}_{L}+n\vec{\beta}_{0}$,  these non-diagonal vertices obviously satisfy GSO equations. 
Adding  superpartners we generate new states, twisted by
$\vec{w}=2n\vec{\beta}_{0}$. The deformations generated by  
$\vec{\beta}_{1}$,...,$\vec{\beta}_{7}$ are also  consistent with SUSY action. 
It can be shown that the orbifold by the the minimal admissible group appears in the compact sector by this way.

The result of these operation leads to the fulfillment of the requirement of space-time supersymmetry.

It is easy to check the mutual locality of any pair of the full vertices obtained by this way. Indeed the monodromy phase is given by
\begin{equation}
	\begin{aligned}
		&(\vec{\mu}_{L}-m^{i}\vec{\beta}_{i})\cdot
		(\tilde{\vec{\mu}}_{L}-\tilde{m}^{j}\vec{\beta}_{j})-\vec{\mu}_{R}\cdot\vec{\mu}_{R}=
		\\
		&\tilde{m}^{i}m^{j}\vec{\beta}_{i}\cdot\vec{\beta}_{j}-\tilde{m}^{j}\vec{\beta}_{j}\cdot\vec{\mu}_{L}-m^{i}\vec{\beta}_{i}\cdot\tilde{\vec{\mu}}_{L}\in\mathbb{Z}.
	\end{aligned}
\end{equation}	

\section{Mutual locality of full physical vertices in orbifolds.}
\label{sec:7}
In principle, it is possible to construct more models starting from
a product $N=2$ minimal models with the total central charge $c=9$	
\begin{equation}
	\begin{aligned}
		&M_{\vec{k}}=\prod_{i=1}^{5}M_{k_{i}}, \ c_{i}=\frac{3k_{i}}{k_{i}+2}, \ c=\sum_{i}c_{i}=9.
	\end{aligned}
\end{equation}
Such possibilities arise if the so-called admissible group of the product of minimal models $\prod_{i=1}^{5}M_{k_{i}}$ includes elements other than the selected element
$(1,1,1,...,1)$.

By the definition the admissible group is any subgroup of
$\mathbb{Z}_{k_{1}+2}\times...\times\mathbb{Z}_{k_{5}+2}$, whose generators include the distinguished element
$(1,1,1,...,1)$, as well as  the elements
$\vec{w}=\{w_1,w_2,...,w_5\}$, which satisfy the following relations
\be
\sum_{i} \frac{w_i}{(k_i+2)}\in\mathbb{Z}.
\label{6.Gadm}
\en 	

To construct more general (orbifold) compactifications we should consider the twisted vertices, with the shifts $\vec{\mu}_{L}\rightarrow\vec{\mu}_{L}+\vec{\nu}$, where
\begin{equation}
\vec{\nu}=(0,0,-2\vec{w},\vec{0}),
\end{equation}
{where $\vec{w}\in G_{adm}\subset G^{max}_{adm}$.
Such deformation satisfies the GSO condition.}
{Consider an arbitrary pair of diagonal mutually local vertices  and twist them by the elements ${\vec{w}}$ and $\tilde{\vec{w}}$:}
\begin{equation}
\vec{\mu}_{L}\rightarrow \vec{\mu}_{L}+\vec{\nu}_{\vec{w}}, \
\tilde{\vec{\mu}}_{L}\rightarrow \tilde{\vec{\mu}}_{L}+\vec{\nu}_{\tilde{\vec{w}}}.
\end{equation}
The monodromy phase for this pair of vertices is given by
\begin{equation}
\begin{aligned}
\exp{[\im 2\pi\sum_{i}\frac{\tilde{w}^{i}(q^{i}_{L}-w^{i})+w^{i}(\tilde{q}^{i}_{L}-\tilde{w}^{i})}{k_{i}+2}]}.
\end{aligned}
\end{equation}

Thus in order to be mutually local the  vertices from  each $\vec{w}$ twisted sector must satisfy 
the following equations
\begin{equation}
\sum_{i=1}^{5}\frac{\gamma_{\tau}^{i}(q^{i}_{R}-w^{i})}{k_{i}+2}\in\mathbb{Z},
\label{Loc}
\end{equation}
where $\vec{\gamma}_{\tau}$ is any  $G_{adm}$ generator.

It can be shown that applying SUSY operators to this mutually local subset of vertices does not violate mutual locality. In this way we generate supermultiplets of mutually local complete vertices. Due to the right "GSO" conditions this set of vertices is also compatible  with the action  $E(8)\times E(6)$ gauge group.

\section{Massless vertices in an explicit form.}



For phenomenological applications, the most important are the massless states of the heterotic string, compactified to four dimensions.

In this section, we explicitly present complete vertices for massless physical states. We omit expressions for the vertices of the graviton, dilaton, and gauge supermultiplets, which are universal and do not depend on the compact factor  and  consider only  construction of  matter massless supermultiplets  which are related with the properties of the compact factor.

\subsection{ $27$ and $\bar{27}$ $E(6)$ supermultiplets}
It can be shown that BRST invariant vertices of the massless scalar   supermultiplets covariant under spacetime supersymmetry  must be chiral or anti-chiral primary states \cite {LVW} in the left-moving sector, such that
\begin{equation}
Q_{L}=\pm 1, \ \Delta_{L,int}=\fr12 |Q_{L}|.
\label{7.ScalL}
\end{equation}
The total left dimension of a vertex is given by
\begin{equation}
\Delta_{L}=-\frac{q}{2}(q+2)+\frac{\lm^{2}}{2}+\Delta_{L,int}=1,
\end{equation}
therefore, in the picture $q=-1$ for the massless scalar particle we  find $\Delta_{L,int}=\frac{1}{2}$. 

On the other hand, the GSO equation requires $Q_{L}=1+ 2\mathbb {Z}$,
as shown above. Taking into account also that the compact factor is the unitary  $N=2$  SCFT, 
we get the  inequality  
\begin{equation}
\Delta_{L,int}\geq\fr12|Q_{L}|
\label{7.UnitL}
\end{equation}
and see that the equality is reached  for the chiral or anti-chiral primaries. As a result   for the left vertices we have
\begin{equation}
\vec{\mu}_{L}=(-1,0,0,\vec{q}_{L}=\pm\vec{l},\vec{s}_{L}=0), \ \sum_{i}\frac{l_{i}}{k_{i}+2}=1.
\end{equation}
As for the right factors, they should be massless and $BRST$-invariant vertices with conformal dimension $\Delta_{R}=1$.  The right vertices  are also characterized by their $E(8)$-weights $\vec{\eps}$, $SO(10)$-weights $\vec{\Lm}$ and $U(1)$-charge $ Q_{R}$.

It can be shown that the right-hand "GSO" equation (\ref{5.GSO}) requires the formation the set of weight vectors of the algebra $E( 6)$ from weights $\vec{\Lm}$ and from the total $U(1)$ charge $Q_{R}$.  This set includes the weight vector  $(\vec { \Lm},\frac{1}{\sqrt{3}}\sum_{i}Q^{i}_{R})$.

What about  the gauge symmetry $E(8)$, then for it these vertices must be singlets. 
Thus, the right-moving factors of massless vertex is determined by
\begin{equation}
\vec{\mu}_{R}=(\vec{0},\vec{\Lm},\vec{q}_{R},\vec{s}_{R})
\end{equation}
The conformal dimensions of such state is
\begin{equation}
\Delta_{R}=\frac{\vec{\Lm}^{2}}{2}+\Delta_{R,int}
\end{equation}

Since in the compact factor we have unitary $N=2$ SCFT the inequality similar to (\ref{7.UnitL}) takes place for the right-moving sector also
\begin{equation}
\Delta_{R,int}\geq\fr12|Q_{R}|.
\label{7.UnitR}
\end{equation}

Now it is easy to see that only two $E(6)$ multiplets gives the massless vertices.
The first  of  massless matter multiplets  is in the $27$ representation of  $E(6)$ with the highest weight vector
$(\vec{\Lm},\frac{1}{\sqrt{3}}\sum_{i}Q^{i}_{R})=(\vec{0},\frac{2}{ \sqrt{3}})$. So its total $U(1)$ charge is equal to 2 and $\Delta_{R,int}=1$. 
 That is
\begin{equation}
\vec{\mu}_{R}=(\vec{\eps}=0,\vec{\Lm}=0,\vec{q}_{R},\vec{s}_{R}=0), \ \sum_{i}\frac{q^{i}_{R}}{k_{i}+2}=2.
\label{8.27}
\end{equation}

The second  massless matter multiplet is given by the $E(6)$ fundamental weight  $e_{1}+\frac{1}{\sqrt{3}}e_{6}$ which is conjugated to the fundamental weight $(0,\frac{2}{\sqrt{3}})$. This representation is $\bar{27}$ of $E(6)$. In this case $\frac{\Lm^{2}}{2}=\frac{1}{2}$ and $\Delta_{R,int}=1$. Therefore it must be chiral primary in the right-moving sector. Hence,
\begin{equation}
\vec{\mu}_{R}=(\vec{\eps}=0,\vec{\Lm}=\vec{v},\vec{q}_{R},\vec{s}_{R}=0), \ \sum_{i}\frac{q^{i}_{R}}{k_{i}+2}=1.
\label{8.27bar}
\end{equation}

For the case of compactification on the model $M_{\vec{k}}=\prod_{i=1}^{5}M_{3}$, factorized by the minimal admissible group $G^{0}_{adm }$, generated by the vector $ \gamma_{0}=(1,1,1,1,1)$, we find only one $27$-multiplet.

Indeed, it is easy to see that there are 101 left-moving chiral primary states with the charge $Q_{L}=1$. So the problem is to find complete vertices, whose left-moving factor is given by one of these $101$ chiral-primaries while they are charged $Q_{R}=2$ chiral primary in the right-moving sector. 
Since $Q_{L}\neq Q_{R}$ these vertices can only appear in the twisted sectors generated by the spectral flow operators $\bar{U}^{\vec{w}}$, which we assume for convenience act in the right-moving sector and where $\vec{w}=n\gamma_{0}$.

Using the explicit expressions for the chiral-chiral and chiral-anti-chiral primary fields in the orbifolds \cite{BP}, see also  \cite{BBP}, \cite {P}, we realize that among the mutually local fields the only vertex of this kind appears in the twisted sector $\vec{w}=2\gamma_{0}$. This vertex can be represented as 
\begin{equation}
	\begin{aligned}
		&\exp{[-\phi + \im p_{\mu}X^{\mu}(z,\bar{z})]}\Phi^{\vec{l}}_{\vec{l}}(z)
		\times\bar{\Phi}^{\tilde{\vec{l}}}_{\tilde{\vec{l}}}(\bar{z}),
		\
		l_{i}=1,\ \tilde{l}_{i}=2,
		\\
		& + \ (10 + 16)\ E(6)\,\, \text{partners} + \text{superpartners}.
		\label{7.27}
	\end{aligned}
\end{equation}

Similarly we find $\bar{27}$ massless matter supermultiplets.  This time one has to find the full vertices, whose compact factor is given by one of the $101$ chiral-chiral-primaries with charges $Q_{L}=Q_{R}=1$. It gives $101$ supermultiplet of the form
\begin{equation}
	\begin{aligned}
		&\exp{[-\phi + \im\bar{H}_{1} +\im p_{\mu}X^{\mu}(z,\bar{z})]}\Phi^{\vec{l}}_{\vec{l}}(z)\times
		\bar{\Phi}^{\vec{l}}_{\vec{l}}(\bar{z})
		 \sum_{i}\frac{l_{i}}{k_{i}+2}=1, \ 
		\\
		& + \ (10 + 16)\ E(6)\,\, \text{partners} + \text{superpartners}.		
	\end{aligned}
\end{equation}

\subsection{Massless singlets}
Now consider the second part of massless physical states, which are massless singlets of the algebra $E(8)\times E(6)$.
The singlets  have a weight vector equal to zero
\begin{equation}
	(\vec{\Lm}=0, \frac{1}{\sqrt{3}}\sum_{i}Q^{i}_{R}=0).
	\label{7.Sing1}
\end{equation}
To be massless and $BRST$-invariant, their vertices  must be $N=0$ primary fields of the Virasoro algebra with right-side conformal dimension equal to one. Therefore, among the mutually local fields, we need to find complete vertices that are the products of the left chiral (or antichiral) primary field and the right primary field such that
\begin{equation}
	\vec{\Lm}=0, \ Q_{R}=0, \ \Delta_{R}=1, \ Q_{L}=\pm 1, \ \Delta_{L}=\fr12.
	\label{7.Sing2}
\end{equation}

For the model $M_{\vec{k}}=\prod_{i=1}^{5}M_{3}$ factored by the minimal admissible group $G^{0}_{adm}$ (quintic model) we find 305 states coming from untwisted sector
\begin{equation}
	\begin{aligned}
		&{\cal{W}}_{\vec{\mu}_{L},\vec{\mu}_{R}}(z,\bar{z})=\exp{[-\phi]}\Phi^{\vec{l}}_{\vec{l}}(z)\times \bar{G}^{-}_{i,-\frac{1}{2}}\bar{\Phi}^{\vec{l}}_{\vec{l}}(\bar{z})
		\exp{[\im p_{\mu}X^{\mu}(z,\bar{z})]},
		\\ &\vec{\mu}_{L}=(-1,0,0,\vec{q}_{L}=\vec{l},\vec{s}_{L}=0), \ \vec{\mu}_{R}=(\vec{\eps}=0,\vec{\Lm}=0,\vec{q}_{R}=\vec{l},\vec{s}_{R}), \ \vec{s}_{R}=(0,...,2_{i},...,0),
		\\ 
		&\sum_{j}\frac{l_{j}}{k_{j}+2}=1,
		\\
		&+\text{superpartners}.
		\label{7.305}
	\end{aligned}
\end{equation}
In the twisted sector $\vec{w}=\gamma_{0}$ we obtain 5 states more
\begin{equation}
	\begin{aligned}
     &{\cal{W}}_{\vec{\mu}_{L},\vec{\mu}_{R}}(z,\bar{z})=
		\exp{[-\phi + \im p_{\mu}X^{\mu}(z,\bar{z})]}\Phi^{\vec{l}}_{-\vec{l}}(z)\times \bar{G}^{-}_{i,-\frac{1}{2}}\bar{\Phi}^{\vec{l}}_{\vec{l}}(\bar{z})
     \\ 
     &\vec{\mu}_{L}=(-1,0,0,\vec{q}_{L}=-\vec{l},\vec{s}_{L}=0), \ \vec{\mu}_{R}=(\vec{\eps}=0,\vec{\Lm}=0,\vec{q}_{R}=\vec{l},\vec{s}_{R}), \ \vec{s}_{R}=(0,...,2_{i},...,0),
     \\ 
     &\vec{l}=(1,1,1,1,1)
     \\
     &+\text{superpartners}.
     \label{7.5}
     \end{aligned}
 \end{equation}
Notice that right-moving factors of these fields are not $N=2$ Virasoro superalgebra primaries but $N=0$ Virasoro algebra primaries.

In addition we find 20 states which are $N=2$ Virasoro superalgebra primaries in the right-moving sector
\begin{equation}
	\begin{aligned}
		&{\cal{W}}_{\vec{\mu}_{L},\vec{\mu}_{R}}(z,\bar{z})=
		\exp{[-\phi + \im p_{\mu}X^{\mu}(z,\bar{z})]}\Phi_{\vec{l}_{L},\vec{q}_{L}}(z)\times\bar{\Phi}_{\vec{l}_{R},\vec{q}_{R}}(\bar{z})
		\\
		&\vec{\mu}_{L}=(-1,0,0,\vec{q}_{L}=\vec{l}_{L},\vec{s}_{L}=0),
		\
		\vec{\mu}_{R}=(\vec{\eps}=0,\vec{\Lm}=0,\vec{q}_{R},\vec{s}_{R}=0),
		\\
		&\vec{l}_{L}=(0,1,1,1,2),
		\ \vec{l}_{R}=(3,1,1,1,2), \ \vec{q}_{R}=(3,-1,-1,-1,0),
		\\ 
		&+ \text{permutations}+\text{superpartners}.
		\label{7.20}
	\end{aligned}
\end{equation}

They all appear in the twisted right - moving  sector corresponding  to $\vec{w}=\gamma_{0}$, and have the  net right-moving charge  $Q_{R}=0$. Therefore, we find a total of 330 singlets $E(8)\times E(6)$, which is consistent with \cite{Gep}.

\section{Conclusion}
In this paper we developed a way to explicitly construct the physical vertices in Gepner models of the Heterotic string compactification. We used the requirement for the simultaneous fulfillment of mutual locality of the left-moving vertices with the space-time symmetry generators and of right-moving vertices with
generators of $E(8)\times E(6)$-gauge symmetry together with the requirement of mutual locality of complete (left-right) vertices of physical states. 
Applying the remaining requirements of the conformal bootstrap, we obtain the same set of physical states as with simultaneous fulfillment of space-time symmetry with the requirement of modular invariance. Our approach is extended for the orbifold models compactification as soon as the admissible group is fixed. The explicit expressions for physical states in the orbifold compactifications is essentially based on our previous papers \cite{BP},\cite{BBP},\cite{P}, where we developed the spectral flow construction of states for the orbifolds of products of Minimal models.
The resulting explicit expressions for physical vertices can be used to calculate string amplitudes, taking into account that the models under consideration are exactly solvable.

\section*{Acknowledgments} 
Authors acknowledge D. Gepner, B. Feigin, A. Litvinov and A. Ilichevsky  for helpful and interesting discussions. A. B. is grateful to  Weizmann Institute of Science and  Joseph Meyerhoff Visiting Professorship for the hospitality  in the time while some of this work was done. This work is supported by the Russian Science Foundation grant 23-12-00333.

\end{document}